\documentclass[aps,pra,superscriptaddress,showpacs,twocolumn,10pt]{revtex4-1}
\usepackage{bm}
\usepackage{amsmath}
\usepackage{textcomp}
\allowdisplaybreaks
\usepackage{longtable}
\usepackage{dcolumn}
\newcolumntype{.}{D{x}{}{-1}}

\newcommand*{\centt}[1]{\multicolumn{1}{c}{#1}}
\newcolumntype{w}[1]{D{.}{.}{#1}}

\newcommand{\bsigma}{\vec{\sigma}}

\newcommand{\bfp}{\vec{p}}

\newcommand{\bfr}{\vec{r}}

\newcommand{\Za}{{Z\alpha}}
\newcommand{\MP}{\vec{p}_1\cdot\vec{p}_2}
\newcommand{\bl}{\ln k_0}
\newcommand{\im}{{\dot{\iota}}}

\usepackage{graphicx}
\usepackage{xcolor}

\begin{document}

\title{Quantum-electrodynamic corrections to the $\bm{1s3d}$ states of the helium atom}

\author{Albert Wienczek}
\affiliation{Faculty of Physics, University of Warsaw,
             Pasteura 5, 02-093 Warsaw, Poland}

\author{Krzysztof Pachucki}
\affiliation{Faculty of Physics, University of Warsaw,
             Pasteura 5, 02-093 Warsaw, Poland}

\author{Mariusz Puchalski}
\affiliation{Faculty of Chemistry, Adam Mickiewicz University, Umultowska 89b, 61-614 Pozna{\'n}, Poland}

\author{Vojt\v{e}ch Patk\'o\v{s}}
\affiliation{Faculty of Mathematics and Physics, Charles University,  Ke Karlovu 3, 121 16 Prague
2, Czech Republic}

\author{Vladimir A. Yerokhin}
\affiliation{Center for Advanced Studies, Peter the Great St.~Petersburg Polytechnic University,
Polytekhnicheskaya 29, 195251 St.~Petersburg, Russia}

\begin{abstract}

We perform quantum electrodynamic calculations of the ionization energy of the $1s3d$ states of
the $^4$He atom, including a complete evaluation of the $m\alpha^6$ correction. We find a large
contribution from the nonradiative part of this correction, which has not been accounted for in
previous investigations. The additional contribution shifts theoretical predictions for
ionization energies by about 10$\,\sigma$. Despite this shift, we confirm the previously reported
systematic deviations between measured experimental results and theoretical predictions for
transitions involving $3D$ states. The reason for these deviations remains unknown.

\end{abstract}

\maketitle

A steadily increasing accuracy of spectroscopic experiments on the helium atom opens new
possibilities for improved determinations of fundamental physical constants, tests of the Standard
Model of fundamental interactions and a search for the new physics. The recent measurement of the
$2\,^3\!P_1$--$2\,^3\!P_2$ helium transition frequency with an accuracy of 25~Hz \cite{hessels:18}
demonstrated a potential for determining the fine-structure constant $\alpha$ with a sub-ppb
accuracy. The main obstacle in achieving this goal is that the present theory of the helium fine
structure \cite{Pachucki:11} is not yet developed enough. Another prominent example is the recent
measurement of the $2\,^3\!P$--$2\,^3\!S$ transition frequency with an accuracy of 1.4~kHz
\cite{zheng:17}. This accuracy is sufficient for the determination of the nuclear charge radius
with a precision below 0.1\%, which is better than what is expected from the muonic helium Lamb
shift. This determination also requires further developments of the helium theory, the
corresponding project being underway \cite{yerokhin:18:betherel}.

It has been previously pointed out \cite{pachucki:17} that experimental results for helium
transitions involving $3D$ states do not agree well with theoretical predictions. The theoretical
values of energy levels of the $D$ states were obtained by Drake and co-workers
\cite{drake:87:prl,drake:92,goldman:92,morton:06:cjp} and have not been verified by independent
calculations. Moreover, their calculations did not fully account for the $m\alpha^6$ QED effects,
in contrast to more complete calculations available for the $n = 1$ and $n = 2$ states
\cite{yerokhin:10:helike,pachucki:17}. Motivated by the reported disagreements, in this work we
perform calculations of the ionization energies of the $1s3d$ states of $^4$He. We extend the
previous works \cite{drake:87:prl,drake:92,goldman:92,morton:06:cjp} by completing the leading QED
effects of order $m\alpha^5$ and performing calculations of the next-order corrections of orders
$m\alpha^6$ and $m\alpha^5\,m/M$.

\section{NRQED expansion}
Within the QED theory, the bound-state energies are defined as the positions of the poles of the
Fourier transform of the equal-time $n$-particle propagator as a function of the complex energy
argument. To calculate the position of these poles for light atoms, it is convenient to use the
nonrelativistic QED (NRQED), which is an effective quantum field theory that gives the same
predictions as the full QED in the region of small momenta, {\em i.e.}, those of the order of the
characteristic electron momentum in an atom.

The basic assumption of the NRQED is that the bound-state energy $E$ can be expanded in powers of
the fine-structure constant $\alpha$,
\begin{eqnarray}\label{eq:1}
E\Bigl(\alpha, \frac{m}{M}\Bigr) &=& \alpha^2\,E^{(2)}\Bigl(\frac{m}{M}\Bigr)
+ \alpha^4\,E^{(4)}\Bigl(\frac{m}{M}\Bigr)
+ \alpha^5\,E^{(5)}\Bigl(\frac{m}{M}\Bigr)
 \nonumber \\&&
 + \alpha^6\,E^{(6)}\Bigl(\frac{m}{M}\Bigr)
 + \alpha^7\,E^{(7)}\Bigl(\frac{m}{M}\Bigr)
+\ldots \,,\label{01}
\end{eqnarray}
where $m/M$  is the electron-to-nucleus mass ratio and the expansion coefficients $E^{(n)}$ may
contain finite powers of $\ln\alpha$. The coefficients $E^{(i)}(m/M)$ are further expanded in
powers of $m/M$:
\begin{equation}\label{eq:2}
E^{(i)}\Bigl(\frac{m}{M}\Bigr) = E^{(i,0)} + \frac{m}{M}\,E^{(i,1)} + \Bigl(\frac{m}{M}\Bigr)^2\,E^{(i,2)} + \ldots\,.
\end{equation}
According to NRQED, the expansion coefficients in Eqs.~(\ref{eq:1}) and (\ref{eq:2}) can be
expressed as expectation values of some effective Hamiltonians with the nonrelativistic wave
function. The derivation of these effective Hamiltonians is the central problem of the NRQED
approach. While the leading-order expansion terms are simple, formulas become increasingly
complicated for higher powers of $\alpha$.

\section{Nonrelativistic energy}
The first term of the NRQED expansion of the bound-state energy, $E^{(2,0)} \equiv E$, is the
eigenvalue of the Schr\"odinger-Coulomb Hamiltonian in the infinite nuclear mass
limit,
\begin{equation}
H_0 \equiv H = \frac{p_1^2}{2}+ \frac{p_2^2}{2} - \frac{Z}{r_1} - \frac{Z}{r_2} + \frac{1}{r}\,,
\end{equation}
where $r_i = |\vec r_i\,|$ and $\vec r = \vec r_1 - \vec r_2$. The finite-nuclear-mass corrections
are induced by the nuclear kinetic energy operator $(m/M)\,\delta_M H$,
\begin{equation}
\delta_M H = \frac{\vec{P}^2}{2}\,, \label{06}
\end{equation}
where $-\vec{P}$ is the nuclear momentum, and in the center-of-mass frame  $\vec{P} = \vec p_1+\vec p_2$.
In the literature, $\delta_M H$ is often separated into two parts:
\begin{equation} \label{eq:rm}
  \delta_M H = \frac{p_1^2+p_2^2}{2} + \vec p_1\cdot\vec p_2\,.
\end{equation}
The first part can be absorbed in the nonrelativistic Hamiltonian by introducing the reduced mass,
whereas the second part is called the mass polarization operator. In the present work, we prefer to
express the recoil corrections in terms of $\delta_M H$, since it makes the resulting formulas
simpler and more transparent.

The first- and second-order recoil corrections to the nonrelativistic energy are given by
\begin{align}
E^{(2,1)} =& \langle\delta_M H\rangle\,,\label{04}\\
E^{(2,2)} =& \Bigl\langle\delta_M H\,\frac{1}{(E-H)'}\,\delta_M H\Bigr\rangle\,.\label{05}
\end{align}
It is also possible to account for the nonrelativistic recoil effect nonperturbatively, by
including $(m/M)\,\delta_M H$ into the nonrelativistic Hamiltonian. In the present work, we use the
nonperturbative approach. For the convenience of the presentation, we express the complete
nonrelativistic energy as $E^{(2)} = E^{(2,0)} + E^{(2,1)} + E^{(2,2+)}$, where $E^{(2,2+)}$
contains corrections of second and higher orders in $m/M$.

The spatial part of the nonrelativistic wave function of a $D$ state is represented in Cartesian
coordinates as a second-rank traceless and symmetric tensor $\phi^{ij}$,
\begin{align} \label{eq:phi}
\phi^{ij}(^{1,3}\!D) &\ = (r_1^i\, r_1^j)^{(2)} F
 + ( r_1^i\, r_2^j)^{(2)} G \pm (1\leftrightarrow2)\,,
\end{align}
where $(r_a^i\, r_b^j)^{(2)} = \frac12(r_a^ir_b^j + r_a^jr_b^i - \frac23\,\delta^{ij}r_a^kr_b^k)$
and the upper (lower) sign corresponds to the singlet (triplet) state, respectively. The functions
$F$ and $G$ are scalar functions of $r_1$, $r_2$, and $r$. In our case they are chosen to be linear
combinations of exponentials of the form $e^{-\alpha\,r_1-\beta\,r_2-\gamma\,r}$ with different
nonlinear parameters $\alpha, \beta$, and $\gamma$. The normalization is taken to be
\begin{equation}
  \langle\phi^{ij}|\phi^{ij}\rangle = 1\,.
\end{equation}
Here and in what follows, we assume the implicit summation over the repeated Cartesian indices. The matrix
element of the nonrelativistic Hamiltonian (or any other spin-independent operator) between the
states $a$ and $b$ is of the form
\begin{equation}
\langle a| H|b\rangle =   \langle\phi_a^{ij}|H|\phi_b^{ij}\rangle. \label{scalar}
\end{equation}
The Hamiltonian is represented as a large square matrix, whose eigenvalues are upper bounds of
the exact nonrelativistic energies. By increasing the size of the basis, one determines the
nonrelativistic energy with a well-controlled uncertainty. The obtained nonrelativistic wave
functions are used for calculating relativistic and QED corrections discussed in the next sections.

\section{Leading-order relativistic correction}

The leading relativistic correction to the nonrelativistic energy is of order $m\alpha^4$ and is
given by the expectation value of the Breit Hamiltonian, which is of the form
\begin{align} \label{eq:h4plus}
  H^{(4+)} =&\, Q_A\Big(\frac{m}{M},a_e\Big) + \vec Q_B\Big(\frac{m}{M},a_e\Big)\cdot\frac{(\vec\sigma_1+\vec\sigma_2)}{2}
 \nonumber \\ &\
  + \vec Q_C\Big(\frac{m}{M},a_e\Big)\cdot\frac{(\vec\sigma_1-\vec\sigma_2)}{2}
  + Q_D^{ij}\Big(\frac{m}{M},a_e\Big)\,\sigma_1^i\,\sigma_2^j\,.
\end{align}
The operators $Q_i$ include the dependence on the nuclear mass $M$ and the electron anomalous
magnetic moment (amm) $a_e = \alpha/(2\pi) + \ldots$. They are given by
\begin{align}
Q_{A}\Big(\frac{m}{M},a_e\Big) =&\, -\frac{1}{8}\,(p_1^4+p_2^4)+
\frac{Z\,\pi}{2}\,[\delta^3(r_1)+\delta^3(r_2)]
 \nonumber\\
&\,
+\pi\,\delta^3(r)
-\frac{1}{2}\,p_1^i\,
\biggl(\frac{\delta^{ij}}{r}+\frac{r^i\,r^j}{r^3}\biggr)\,p_2^j\nonumber \\ &\,
-\frac{Z\,m}{2\,M}\,\biggl[
p_1^i\,\biggl(\frac{\delta^{ij}}{r_1} + \frac{r_1^i\,r_1^j}{r_1^3}\biggr)
 \nonumber\\ &\
+
p_2^i\,\biggl(\frac{\delta^{ij}}{r_2} + \frac{r_2^i\,r_2^j}{r_2^3}\biggr)\biggr]\,P^j
\,,\\
\vec Q_B\Big(\frac{m}{M},a_e\Big) =&\,\frac{Z}{4}\Big( \frac{\bfr_1}{r_1^3}\times\bfp_1 + \frac{\bfr_2}{r_2^3}\times\bfp_2\Bigr)\,\big(1+2\,a_e\big)
\nonumber \\ &\,
- \frac34 \frac{\bfr}{r^3}\times(\bfp_1-\bfp_2)\,\Big(1+\frac{4}{3}\,a_e\Big)\nonumber \\ &\,
+ \frac{m}{M}\frac{Z}{2}\Big(\frac{\bfr_1}{r_1^3}+\frac{\bfr_2}{r_2^3}\Big)\times \vec P
  \big(1+a_e\big)
\,, \\
\vec Q_C\Big(\frac{m}{M},a_e\Big) =&\, \frac{Z}{4}\Big( \frac{\bfr_1}{r_1^3}\times\bfp_1 - \frac{\bfr_2}{r_2^3}\times\bfp_2\Bigr)
    \big(1+2\,a_e\big)
  \nonumber \\ &\,
+ \frac{1}{4}\,\frac{\bfr}{r^3}\times\vec P
  \nonumber \\ &\,
+ \frac{m}{M}\frac{Z}{2}\Big(\frac{\bfr_1}{r_1^3}-\frac{\bfr_2}{r_2^3}\Big)\times \vec P
  \big(1+a_e\big)  \,, \\
Q_D^{ij}\Big(\frac{m}{M},a_e\Big) =&\,\frac14 \biggl( \frac{\delta^{ij}}{r^3} - 3 \frac{r^ir^j}{r^5}\biggr)\,(1+a_e)^2\,.
\end{align}
The upper index in $H^{(4+)}$ indicates that this Hamiltonian includes operators of order
$m\alpha^4$ and higher (due to the presence of $a_e$ and $m/M$). We also need the Hamiltonian
that contains only $m\alpha^4$ operators, which is obtained from the above equations by setting
$a_e\to 0$ and $m/M\to 0$,
\begin{align}
  H^{(4)} =&\, Q_A + \vec Q_B\cdot\frac{(\vec\sigma_1+\vec\sigma_2)}{2}
  + \vec Q_C\cdot\frac{(\vec\sigma_1-\vec\sigma_2)}{2} + Q_D^{ij}\,\sigma_1^i\,\sigma_2^j\,,
\end{align}
where we assume the short-hand notations $Q_i \equiv Q_i(0,0)$.

The relativistic corrections to the nonrelativistic energy are given by
\begin{eqnarray}
E^{(4,0)} &=&\langle H^{(4)}\rangle \,,
\end{eqnarray}
\begin{eqnarray}
E^{(4,1)} &=& 2\,\langle H^{(4)}\,\frac{1}{(E-H)'}\,\delta_M H\rangle
+ \langle\delta_M H^{(4)}\rangle\,,
\end{eqnarray}
where $\delta_M H^{(4)}$ is the $M$-dependent part of $H^{(4+)}$ (with $a_e\to 0$). The
higher-order (in the mass ratio) terms can be neglected for the $D$ states.

In practical calculations of $E^{(4,0)}$ it is convenient to use instead of $Q_A$ its regularized form of $Q_{A{\rm
reg}}$, given by Eq.~(\ref{eq:QA}), which has the same expectation value on eigenstates of the
(nonrecoil) nonrelativistic Hamiltonian.

The expectation values of spin-dependent operators on the eigenstates of $J^2$ and $J_z$ ($\vec
J=\vec L+\vec S$, where $\vec S = \vec s_1+\vec s_2$) are calculated with help of the following
formulas
\begin{align} \label{fs3}
  \langle {}^3\!D_J|\vec Q\cdot\vec \sigma_a\,|{}^3\!D_J\rangle =&\,  \langle {}^3\!D_J|\vec Q\cdot\vec S\,|{}^3\!D_J\rangle
  \nonumber \\ =&\
      u_J\,\im\,\epsilon^{jli}\,\langle {}^3\!D^{jk} |Q^l|{}^3\!D^{ik}\rangle\,,\\
  \langle {}^3\!D_J|Q^{ij}\,\sigma_1^i\,\sigma_2^j\,|{}^3\!D_J\rangle =&\,  2\,\langle {}^3\!D_J|Q^{ij}\,S^i\,S^j\,|{}^3\!D_J\rangle
  \nonumber \\ =&\, 2\,v_J\,\langle {}^3\!D^{ik}|Q^{ij}|{}^3\!D^{jk}\rangle\,,
   \label{fs3a}
\end{align}
where $Q^i$ is an arbitrary  vector, $Q^{ij}$ is an arbitrary symmetric and traceless tensor
operator, $|{}^3\!D^{ik}\rangle \equiv |\phi^{ik}({}^3\!D)\rangle $ is the spacial part of the
wave function (\ref{eq:phi}), and
\begin{eqnarray}
u_J &=& (1,1/3,-2/3),\\
v_J &=& (-1,1,-2/7)\,,
\end{eqnarray}
for $J=1,2,3$, respectively. The above formulas were derived by
taking into account that
\begin{align}
  \langle {}D_J| Q |{} D_J\rangle =&\,  \frac1{2J+1} \sum_{M_J} \langle {} D_{JM_J}| Q |{} D_{JM_J}\rangle
 \nonumber \\
  =&\, \frac1{2J+1} \sum_{M_J} {\rm Tr} \Big[ Q\, \big|{} D_{JM_J}\big> \big<{} D_{JM_J}\big| \Big]\,,
\end{align}
(where $M_J = -J,\ldots, J$ is the angular momentum projection) and then evaluating traces with help of Eqs.~(\ref{tr1})-(\ref{tr2}).

\section{Leading-order QED}

The leading QED correction to energy levels is of order $m\alpha^5$ and can be expressed by
\begin{eqnarray}
E^{(5,0)} &=&\langle H^{(5)}\rangle \,,
\end{eqnarray}
\begin{eqnarray}
E^{(5,1)} &=& 2\,\langle H^{(5)}\,\frac{1}{(E-H)'}\,\delta_M H\rangle
+ \langle\delta_M H^{(5)}\rangle\,.
\end{eqnarray}
The effective $m\alpha^5$ Hamiltonian is \cite{araki:57,sucher:57}
\begin{eqnarray}
H^{(5)} &=&
\sum_a\left(\frac{19}{30}+\ln(\alpha^{-2}) - \bl \right)\,
\frac{4\,Z}{3}\,\delta^3(r_a)
 \nonumber \\ &&
+ \left(\frac{164}{15}+\frac{14}{3}\,\ln\alpha \right)\,\delta^3(r)
-\frac{7}{6\,\pi}\,\left(\frac{1}{r^3}\right)_{\!\epsilon}
 \nonumber \\ &&
 + H^{(5)}_{\rm fs}
\,,
\end{eqnarray}
where the index $a = 1,2$ numerates the electrons and the spin-dependent operator $H^{(5)}_{\rm
fs}$ is the nuclear-mass-independent $m\alpha^5$ part of $H^{(4+)}$ in Eq.~(\ref{eq:h4plus}).
Further notations are as follows: $\bl$ is the Bethe logarithm defined as
\begin{eqnarray}
\bl &=&
\frac{\big \langle \sum_a\vec p_a \,(H-E)\,\ln \big[2\,(H-E)\big]\,
\sum_b\vec p_b \big\rangle}{2\,\pi\,Z\,\big\langle\sum_c \delta^3(r_{\rm c})\big\rangle}\,,
\end{eqnarray}
 and $\big(1/r^3\big)_{\!\epsilon}$ is the so-called Araki-Sucher term, defined by its matrix
 elements as
\begin{eqnarray}
\biggl\langle\frac{1}{r^3}\biggr\rangle_{\!\epsilon} &=&
\lim_{\epsilon\rightarrow 0}\int {\rm d}^3 r\,
\phi^*(\vec r)\biggl[\frac{1}{r^3}\,\Theta(r-\epsilon)
+ 4\,\pi\,\delta^3(r)\,
  \nonumber \\ && \times
(\gamma+\ln \epsilon)\biggr]\,\phi(\vec r)\,.
\end{eqnarray}

The recoil addition to the $m\alpha^5$ Hamiltonian is given by \cite{pachucki:00:herec}
\begin{eqnarray}
\delta_M H^{(5)} &=&
\sum_a\biggl[\biggl(\frac{62}{3}+\ln(\alpha^{-2})
-8\,\bl - \frac{4}{Z}\,\delta_M\bl \biggr)\,
 \nonumber \\ && \times
\frac{Z^2}{3}\,\delta^3(r_a) -\frac{7\,Z^2}{6\,\pi}\,\left(\frac{1}{r_{a}^3}\right)_{\!\epsilon}\biggr]
+ \delta_M H^{(5)}_{\rm fs} \,,\label{24}
\end{eqnarray}
where $\delta_M H^{(5)}_{\rm fs}$ is the nuclear-mass-dependent $m\alpha^5$ part of  $H^{(4+)}$ in
Eq.~(\ref{eq:h4plus}) and    $\delta_M \ln k_0$ is the correction to the Bethe logarithm $\ln k_0$
induced by the nonrelativistic kinetic energy operator $\delta_M H$ in Eq.~(\ref{06}).

In numerical calculations, it is sometimes convenient to separate $\delta_M H$ into the reduced-mass
and mass-polarization parts according to Eq.~(\ref{eq:rm}). The former can be parametrized
analytically by introducing the reduced mass, whereas the latter needs to be calculated
numerically. The separation of Eq.~(\ref{eq:rm}) leads to
\begin{align}
  \delta_M \left<\frac{1}{r^3}\right>_{\!\epsilon} =&\ \delta_{p_1p_2} \left<\frac{1}{r^3}\right>_{\!\epsilon}
-3\,\left<\frac{1}{r^3}\right>_{\!\epsilon} + \big< 4\,\pi\delta^3(r) \big>\,,\\
  \delta_M\bl =&\ \delta_{p_1p_2}\bl + 1\,,
\end{align}
where $\delta_{p_1p_2}$ denotes the perturbation due to the mass polarization operator $\vec
p_1\cdot\vec p_2$.

In this work we performed direct numerical calculations of the Bethe logarithm for the $1s3d$
states, with the method described in Ref.~\cite{yerokhin:10:helike}. Our numerical results are
presented in Table~\ref{tab:bethe}. They are in good agreement with previous results
\cite{drake:01} obtained by the numerical method developed by Drake and Goldman
\cite{drake:99:cjp}. We also performed calculations of the Bethe logarithm with the mass
polarization term included into the Hamiltonian. We found that the mass polarization contribution
to the Bethe logarithm is very small and cannot be clearly identified at the level of our present
numerical accuracy of a few parts in $10^{-9}$.

\begin{table}
\caption{Numerical results for the Bethe logarithm $\beta \equiv \ln(k_0/Z^2)$ for the $3^1\!D$ and $3^3\!D$ states of helium.
$\delta \beta = \beta -\beta_{1s}$, where
$\beta_{1s}$ is the Bethe logarithm for the hydrogenic $1s$ state, $\beta_{1s} =
2.984\,128\,555\,765\,498$~\cite{drake:90}. For each state, the upper line presents
results for the infinitely massive nucleus; the lower line presents results with inclusion of the
mass polarization contribution $(m/M)\,\MP$. \label{tab:bethe} }
\begin{ruledtabular}
\begin{tabular}{lllll}
  \multicolumn{1}{c}{State} && \multicolumn{1}{c}{$\beta$} & \multicolumn{1}{c}{$\delta \beta\times 10^6$}
                                                          & \multicolumn{1}{c}{$\delta \beta\times 10^6$}\\
                            && & & Ref.~\cite{drake:01} \\
\hline\\[-5pt]
$3^1\!D$ &&   $2.984\,119\,109\,(7)$	   & $-9.447\,(7)$	 &	    $-9.38\,(7)$  \\
&$\MP$    &   $2.984\,119\,106\,(7)$       & $-9.449\,(7)$                  \\[2pt]
$3^3\!D$ &&   $2.984\,125\,886\,(2)$       & $-2.670\,(2)$&          $-2.64\,(11)$ \\
&$\MP$    &   $2.984\,125\,891\,(2)$       & $-2.665\,(2)$&               \\[2pt]
\end{tabular}
\end{ruledtabular}
\end{table}

\section{Singlet-triplet mixing}
\label{sec:mix}

The correction due to mixing of the $3^1\!D$ and $3^3\!D$ states is formally of order $m\alpha^6$
but it is strongly enhanced due to a small energy difference between these states. For this reason
we consider this contribution separately.

We calculate the mixing correction by forming an effective Hamiltonian matrix in the subspace of
the two strongly mixing states,
\begin{align}\label{eq:ham}
  H_{\rm eff} = \left(\begin{array}{cc}
    E_{\rm dia}(^3\!D_2) & E_{\rm off}\\
    E_{\rm off} & E_{\rm dia}(^1\!D_2)
  \end{array}\right)\,,
\end{align}
where
\begin{align}
  E_{\rm dia} =&\ m\alpha^2\,\big[E^{(2)} + \alpha^2\,E^{(4)} + \alpha^3\,E^{(5)}\big],\\
  E_{\rm off} =&\ m\alpha^4\,\langle{}^3\!D_{2M_J}|H_C|{}^1\!D_{1M_J}\rangle\nonumber \\
            =&\ m\alpha^4\,\langle{}^1\!D_{2M_J}|H_C|{}^3\!D_{1M_J}\rangle\,,
\end{align}
and $H_C =\vec{Q}_C(a_e,m/M)\,\cdot(\vec \sigma_1-\vec\sigma_2)/2$ is the part of the Breit
Hamiltonian $H^{(4+)}$ that mixes the triplet and singlet states. The mixing correction is obtained
by diagonalizing the effective Hamiltonian (\ref{eq:ham}), with the result
\begin{align}
E_{\rm MIX}({}^1\!D_2) = -E_{\rm MIX}({}^3\!D_2) =\frac12\sqrt{(\Delta E)^2+4 E_{\rm off}^2} -\frac12 \Delta E\,
\end{align}
where $\Delta E = E_{\rm dia}(3{}^1\!D_2) - E_{\rm dia}(3{}^3\!D_2) > 0$ and the square of
the off-diagonal term is evaluated as
\begin{align}
  |\langle{}^3\!D_{2M_J}|H_C|{}^1\!D_{1M_J}\rangle|^2 =
 \frac23\,\big< {}^3\!D^{ik} \big| \epsilon^{klj}\,\im\,Q_C^l\big|{}^1\!D^{ij}\big\rangle^2\,.
\end{align}

\begin{widetext}

\section{$\bm{m\alpha^6}$ QED}

The $m\alpha^6$ correction to the energy levels was derived in
Refs.~\cite{hsinglet,hsinglet2}. It can be represented as a sum of the first-order and second-order
perturbation corrections induced by various effective Hamiltonians,
\begin{align}
E^{(6)} = E_Q + E_H + E_{R1} + E_{R2} + E_{LG}  + E_{\rm fs, DK} + E_{\rm fs, amm} + E_{\rm sec}\,,
\end{align}
where
\begin{align}
E_Q &\ = \Big< -\frac{E^3}{2}-\frac{1}{8} E Z Q_{1}+\frac{Q_{2}}{8}+\frac{1}{8} Z(1-2Z) Q_{3}+\frac{3}{16} Z Q_{4}-\frac{1}{4} ZQ_{5}
+\frac{Q_{6S}}{24}-\frac{({\cal S}+3)}{96} Q_{6T}\nonumber\\
&
+\frac{1}{4} \big(E^2+2E^{(4,0)}\big) Q_{7}-\frac{(5 {\cal S}+31)}{32} E Q_{8} +\frac{(5 {\cal S}+23)}{32} Q_{9}
+\frac{1}{2}   E Z^2Q_{11} +E Z^2Q_{12}
\nonumber\\
&
-E ZQ_{13} -Z^2Q_{14} + Z^3Q_{15} -\frac{1}{2} Z^2Q_{16} -\frac{(5 {\cal S}+23)Z}{16} Q_{17}-\frac{(5{\cal S}+13) Z}{32} Q_{18}
 \nonumber\\
&
+\frac{1}{2} Z Q_{19}-\frac{1}{8} Z^2Q_{20}   +\frac{1}{4} Z^2 Q_{21} +\frac{1}{4}Z^2 Q_{22}
+\frac{(5 {\cal S}+47)}{32} Q_{23} +\frac{1}{2}Z Q_{24}+\frac{({\cal S}-3)}{192}Q_{25}
\nonumber\\
&
-\frac{1}{4}Z Q_{26} -\frac{1}{8}   E Q_{27}
-\frac{1}{2}Z Q_{28}+\frac{Q_{29}}{4}+\frac{Q_{30}}{8}\Big>\,,
\end{align}
where ${\cal S} = \vec{\sigma}_1\cdot\vec{\sigma}_2$, $\big< {\cal S}\big> = -3$ for singlet and
$\big< {\cal S}\big> = 1$ for triplet states, and operators $Q_i$ are defined in Table~\ref{tab:Q}.
$E_H$ is the high-energy contribution induced by the forward three-photon exchange scattering amplitude,
\begin{align}
  E_{H} = \bigg[ -\frac{39 \zeta(3)}{\pi^2}+\frac{32}{\pi^2} - 6\ln 2 + \frac73\bigg] \frac{\pi}{4}\big< \delta(r)\big>\,,
\end{align}
and $E_{LG}$ is the logarithmic contribution,
\begin{align}
E_{LG} = -\pi \ln \alpha \big< \delta(r)\big>\,.
\end{align}

$E_{R1}$ and $E_{R2}$ are the radiative one-loop and two-loop contributions, respectively,
\begin{align}
  E_{R1} &\ = Z^2\,\biggl[\frac{427}{96}-2\,\ln2\biggr]\,\pi\,
  \bigl<\delta^3(r_1)+\delta^3(r_2)\bigr>
  +\biggl[
  \frac{6\,\zeta(3)}{\pi^2}-\frac{697}{27\,\pi^2}-8\,\ln2+\frac{1099}{72}
  \biggr]\,\pi\,\big<\delta^3(r)\big>,
\end{align}
\begin{align}
  E_{R2} &\ =
  Z\,\biggl[-\frac{9\,\zeta(3)}{4\,\pi^2}-\frac{2179}{648\,\pi^2}+
  \frac{3\,\ln2}{2}-\frac{10}{27}\biggr]\,\pi\,
  \bigl<\delta^3(r_1)+\delta^3(r_2)\bigr>
  +\biggl[\frac{15\,\zeta(3)}{2\,\pi^2}+\frac{631}{54\,\pi^2}
  -5\,\ln2+\frac{29}{27} \biggr]\,\pi\,\big<\delta^3(r)\big>\,.
\end{align}

$E_{\rm fs, DK}$ is the Douglas-Kroll correction to the fine structure,
\begin{align}
  E_{\rm fs, DK}  =& -\frac{3\,Z}{8}\, R_1 -Z\,R_2 + \frac{Z}{2}\,R_3 + \frac{1}{2}\,R_4 - \frac{1}{2}\,R_5
  + \frac{5}{8}\,R_6
  -\frac{3}{4}\,R_7 -\frac{1}{4}\,R_8 - \frac{3}{4}\,R_9 +\frac{3}{8}\,R_{10}\nonumber \\ &\
   -\frac{3}{16}\,R_{11} -\frac{1}{16}\,R_{12}
  +\frac{3}{2}\,R_{13} -\frac{1}{4}\,R_{14} +\frac{1}{8}\,R_{15}\,,
\end{align}
\end{widetext}
where $R_i$ are defined in
Table~\ref{tab:R}, and $E_{\rm fs, amm}$ is the amm correction to the fine
structure, which is the $m\alpha^6$ part of the Breit Hamiltonian $H^{(4+)}$ in
Eq.~(\ref{eq:h4plus}).

\begin{table*}
\caption{Expectation values of operators $Q_i$ with $i=1,\ldots,30$ for the $3{}^1\!D$ and $3{}^3\!D$ states,
$\vec p = (\vec p_1-\vec p_2)/2$, $\vec P = \vec p_1+\vec p_2$. \label{tab:Q}}
\label{oprsQ}
\begin{ruledtabular}
\begin{tabular}{lldd}
            &                                                             &    \multicolumn{1}{c}{$3^1\!D$}
                                                                         &  \multicolumn{1}{c}{$3^3\!D$}  \\ \hline
                                                                          \\[-5pt]
$Q_1 $ & $4 \pi \delta^3 (r_1)$   				                          &  15.99824880    &  15.99784093     \\
$Q_2 $ & $4 \pi \delta^3 (r)$               	                          &   0.00002874    &   0 \\
$Q_3 $ & $4 \pi \delta^3(r_1)/r_2$                  	                  &   1.78124709    &   1.78249111     \\	
$Q_4 $ & $4 \pi \delta^3(r_1)\, p_2^2$ 	                                  &   1.78492102    &   1.78746694     \\
$Q_5 $ & $4 \pi \delta^3(r)/r_1$				                          &   0.00002428    &   0 \\
$Q_{6S} $ & $4 \pi\,\delta^3(r)\,P^2 $				                      &   0.00018470    &   0 \\
$Q_{6T} $ & $4 \pi\,{\vec{p}} \,\delta^3(r)\,\vec p $				          &   0             &   0.00019375 \\
$Q_7 $ & $1/r$						                                      &   0.11121606    &   0.11129738     \\
$Q_8 $ & $1/r^2$						                                  &   0.01496643    &   0.01497321  \\
$Q_9 $ & $1/r^3$                    	                                  &   0.00258825    &   0.00257961  \\
$Q_{10}$ & $1/r^4$                  	                                  &   0.00058978    &   0.00065795  \\
$Q_{11}$ & $1/r_1^2$                	                                  &   4.00705950    &   4.00698329     \\
$Q_{12}$ & $1/(r_1 r_2)$            	                                  &   0.22267879    &   0.22284220     \\
$Q_{13}$ & $1/(r_1 r)$              	                                  &   0.11870346    &   0.11879759     \\
$Q_{14}$ & $1/(r_1 r_2 r)$          	                                  &   0.02973688    &   0.02979293  \\
$Q_{15}$ & $1/(r_1^2 r_2)$					                              &   0.46021055    &   0.46055487     \\
$Q_{16}$ & $1/(r_1^2 r)$					                              &   0.44649022    &   0.44680978     \\
$Q_{17}$ & $1/(r_1 r^2)$   					                              &   0.01618262    &   0.01620066  \\
$Q_{18}$ & $(\vec{r}_1\cdot\vec r)/(r_1^3 r^3)$                           &   0.00039491    &   0.00039799  \\
$Q_{19}$ & $(\vec{r}_1\cdot\vec r)/(r_1^3 r^2)$                           &   0.00678471    &   0.00679459  \\
$Q_{20}$ & $r_1^i r_2^j(r^i r^j-3\delta^{ij}r^2)/(r_1^3 r_2^3 r)$         &   0.00244330    &   0.00231961  \\
$Q_{21}$ & $p_2^2/r_1^2$					                              &   0.47602349    &   0.47671633     \\
$Q_{22}$ & $\vec{p}_1 (1/r_1^2)\, \vec{p}_1$			                      &  16.00022730    &  15.99977930     \\
$Q_{23}$ & $\vec{p}_1(1/r^2)\, \vec{p}_1$			                          &   0.03192362    &   0.03210149  \\
$Q_{24}$ & $p_1^i\,(r^i r^j+\delta^{ij} r^2)/(r_1 r^3)\, p_2^j$           &  -0.00025389    &  -0.00024600  \\
$Q_{25}$ & $P^i\, (3 r^i r^j-\delta^{ij} r^2) (1/r^5)\, P^j$	              &  -0.00054572    &  -0.00051388  \\
$Q_{26}$ & $p_2^k \,r_1^i\,(1/r_1^3) (\delta^{jk} r^i/r - \delta^{ik} r^j/r
-\delta^{ij} r^k/r -r^i r^j r^k/r^3)\, p_2^j$		                  &   0.00081210    &   0.00036806  \\
$Q_{27}$ & $p_1^2\, p_2^2$					                              &   0.44630062    &   0.44689565     \\
$Q_{28}$ & $p_1^2\,(1/r_1)\, p_2^2$				                          &   1.37356243    &   1.37554162     \\
$Q_{29}$ & $\vec{p}_1\times\vec{p}_2\,(1/r)\,\vec{p}_1\times\vec{p}_2$
							                                          &   0.04578880    &   0.04728700  \\
$Q_{30}$ & $p_1^k \,p_2^l\,(-\delta^{jl} r^i r^k/r^3 - \delta^{ik} r^j r^l/r^3
+3r^i r^j r^k r^l/r^5)\, p_1^i\, p_2^j$			                      &  -0.02301349    &  -0.02388848  \\
\end{tabular}
\end{ruledtabular}
\end{table*}

\begin{table*}
\caption{Expectation values of spin-dependent $m\alpha^6$ operators for the $3^3D_J$ states. \label{tab:R}}

\label{oprsR}
\begin{ruledtabular}
\begin{tabular}{lld}
$R_1    $ & $  p_1^2\,(\bfr_1/r_1^3)\times\bfp_1 \cdot \bsigma_1$                                    &   -0.001359311  \  u_J   \\[2pt]
$R_2    $ & $  (\bfr_1/r_1^3)\times(\bfr/r^3)\cdot \bsigma_1\,(\bfr\cdot\bfp_2)$                     &   -0.002475817  \  u_J  \\[2pt]
$R_3    $ & $  (\vec r/r^3)\cdot\vec\sigma_1\,(\vec r_1/r_1^3)\cdot\vec\sigma_2$                     &   -0.000693482  \  2\,v_J   \\[2pt]
$R_4    $ & $  (\bfr/r^4)\times\vec p_2\cdot\vec\sigma_1$                                            &    0.000921565  \  u_J     \\[2pt]
$R_5    $ & $  (\bfr/r^6)\cdot\vec\sigma_1\,\vec r\cdot\vec\sigma_2$                                 &    0.000197304  \  2\,v_J   \\[2pt]
$R_6    $ & $  p_1^2\,(\vec r/r^3)\times\vec p_1\cdot\vec\sigma_1$                                   &   -0.001250197  \  u_J     \\[2pt]
$R_7    $ & $  p_1^2\,(\vec r/r^3)\times\vec p_2\cdot\vec\sigma_1$                                   &    0.014880254  \  u_J     \\[2pt]
$R_8    $ & $  \im\,p_1^2\,(1/r)\,\bsigma_1\cdot(\bfp_1\times\bfp_2)$                                &    0.001726521  \  u_J     \\[2pt]
$R_9    $ & $  \im\,p_1^2\,(\bfr/r^3)\cdot\bfp_2\,\bfr\times\bfp_1\cdot\bsigma_1$                    &   -0.001924622  \  u_J     \\[2pt]
$R_{10} $ & $  \im\,(\vec r/r^5)\times (\vec r\cdot \vec p_2)\,\vec p_1\cdot \vec\sigma_1$           &   -0.000041310  \  u_J     \\[2pt]
$R_{11} $ & $  (\bfr/r^5)\times (\vec r\times \vec p_1\cdot\vec\sigma_1)\,\vec p_2\cdot \vec\sigma_2$&   -0.000088519  \  2\,v_J   \\[2pt]
$R_{12} $ & $  (1/r^3)\,\vec p_1\cdot\vec\sigma_2\,\vec p_2\cdot\vec\sigma_1$                        &    0.000182341  \  2\,v_J   \\[2pt]
$R_{13} $ & $  p_1^2\,(\bfr/r^5)\cdot\vec\sigma_1\,\vec r\cdot\vec\sigma_2$                          &    0.001782698  \  2\,v_J   \\[2pt]
$R_{14} $ & $  \im\,p_1^2\,(\bfr/r^3)\cdot\vec\sigma_1\,\vec p_1\cdot\vec\sigma_2$                   &    0.001084191  \  2\,v_J   \\[2pt]
  $R_{15} $ & $  \im\,p_1^2\,(1/r^3)\,\big[\bfr\cdot\vec\sigma_1\,\vec p_2\cdot\vec\sigma_2 + \bfr\cdot\vec\sigma_2\,\vec p_2\cdot\vec\sigma_1
                -(3/r^2)\,\bfr\cdot\vec\sigma_1\,\bfr\cdot\vec\sigma_2\,(\bfr\cdot\vec p_2)\big]$
                                                                                                   & -0.005402678  \  2\,v_J   \\[2pt]
\end{tabular}
\end{ruledtabular}
\end{table*}

$E_{\rm sec}$ is the second-order correction induced by the Breit Hamiltonian. After the separation
of divergences, it is represented as
\begin{align} \label{eq6:ma6sec}
  E_{\rm sec} =&\  \biggl\langle H^{(4)}_{\rm reg}\,\frac{1}{(E - H)''}\,H^{(4)}_{\rm reg}\biggr\rangle \,,
\end{align}
where $H^{(4)}_{\rm reg}$ is the regularized Breit Hamiltonian defined below. The double prime on
the electron propagator $1/(E-H)''$  indicates that one should exclude from the summation over the
Schr\"odinger spectrum not only the reference state (as is the case in all second-order
corrections), but also the state with the opposite spin coupling. More specifically, for the
$3\,^{2S+1}\!D$ reference states relevant for this work, we exclude {\em both} the $3\,^1\!D$ and
$3\,^3\!D$ states from the summation over the spectrum. We note that the intermediate state with
the opposite spin coupling (triplet for singlet, and vice versa) is already accounted for in the
mixing contribution discussed in Sec.~\ref{sec:mix}.

The regularized Breit Hamiltonian is given by \cite{hsinglet}
\begin{align}
  H^{(4)}_{\rm reg} =&\, Q_{A{\rm reg}} + \vec Q_B\cdot\frac{(\vec\sigma_1+\vec\sigma_2)}{2}
  \nonumber \\ &\
  + \vec Q_C\cdot\frac{(\vec\sigma_1-\vec\sigma_2)}{2} + Q_D^{ij}\,\sigma_1^i\,\sigma_2^j\,,
\end{align}
where
\begin{align} \label{eq:QA}
Q_{A{\rm reg}} =&\ -\frac12\big( E-V\big)^2
- p_1^i \frac1{2r}\Big(\delta^{ij}+\frac{r^ir^j}{r^2}\Big)p_2^j
 \nonumber \\ &
+ \frac14 \vec{\nabla}_1^2  \vec{\nabla}_2^2 -\frac{Z}{4}\Big( {\frac{\vec{r}_1}{r_1^3}\cdot
\vec{\nabla}_1 + \frac{\vec{r}_2}{r_2^3}\cdot \vec{\nabla}_2} \Big) \,,
\end{align}
and $V = -Z/r_1-Z/r_2+1/r$. The operator $\vec{\nabla}_1^2  \vec{\nabla}_2^2$ in the above expression
is non-Hermitian and requires an explicit definition. Its action on a trial function $\phi$ on the
right should be understood as a plain differentiation (omitting $\delta^3(r)$; no differentiation
by parts is allowed in the matrix element). We note that the expectation value of the regularized
Breit Hamiltonian on the eigenfunctions of the (nonrecoil) nonrelativistic Hamiltonian is the same
as that of $H^{(4)}$:
\begin{eqnarray}
\langle H^{(4)}_{\rm reg}\rangle = \langle H^{(4)}\rangle = E^{(4,0)}\,.
\end{eqnarray}

The second-order $m\alpha^6$ correction involves numerous contributions from many different
symmetries of intermediate states. The angular-momentum algebra is performed in Cartesian
coordinates as explained in Appendix~\ref{app:tensor}, with the explicit formulas listed in
Appendix~\ref{app:sec}.

\section{Higher-order QED correction}

We estimate the $m\alpha^7$ correction to the ionization energy of $1snd$ states as
\begin{align}\label{eq:ma7}
E^{(7)} = &\ \bigg[ Z^3 \,\Big(L^2\,A_{62}+L\,A_{61}+ A_{60}\Big)\,
  \nonumber \\ &
 + \frac{Z^2}{\pi}\,B_{50} + \frac{Z}{\pi^2}\,C_{40}\bigg]
\bigg[ \big< \delta(\vec{r}_1)+\delta(\vec{r}_2)\big> - \frac{Z^3}{\pi}\bigg]\,,
\end{align}
where $L = \ln[(Z\alpha)^{-2}]$ and $A_{ij}$, $B_{ij}$, and $C_{ij}$ are the coefficients of the
$\Za$ expansion of one-loop, two-loop, and three-loop QED effects for the $1s$ hydrogenic state,
respectively. The numerical values of the coefficients are $A_{62} = -1$, $A_{61} = 5.286040$,
$A_{60} = -31.501041$, $B_{50}= -21.5544$, and $C_{40} = 0.417504$ \cite{yerokhin:18:hydr}. Having in
mind that in order $m\alpha^6$ the radiative QED correction is one of the largest but not the
dominant contribution, we ascribe the uncertainty of 100\% to this approximation of $E^{(7)}$.

\section{Results and Discussion}

The results of our numerical calculations of the $m\alpha^6$ corrections are listed in
Table~\ref{tab:ma6}. The numerical values presented are corrections to the ionization energy, i.e.,
the corresponding hydrogenic $1s$ contributions are subtracted from $E_Q$, $E_{R1}$, $E_{R2}$, and
$E_{\rm sec}({}^{2S+1}\!D)$. The subtraction of the hydrogenic contribution leads to a cancellation
of about five decimal figures, which makes calculations rather demanding, especially for the $E_{\rm
sec}({}^{2S+1}\!D)$ correction. Specifically, for the $3\,^1\!D_2$ reference state, the numerical
value of $-0.156\,(2)$ quoted in Table~\ref{tab:ma6} for the $E(3\,{}^{1}\!D_2|{}^1\!D)$ arises as
$-16\,000.156\,(2) + 16\,000$, where the latter term is the hydrogenic $1s$ contribution.

The interesting feature about the obtained $m\alpha^6$ results is that the one-loop radiative
correction $E_{R1}$ is {\em not} dominant. The remaining, nonradiative $m\alpha^6$ contribution is
larger than the radiative, and of the opposite sign. As a result, the total $m\alpha^6$ correction
is quite small numerically and differs significantly from the previous estimations
\cite{morton:06:cjp}.
The nonradiative part of $m\alpha^6$ correction, which has not been accounted for in the previous
calculation \cite{morton:06:cjp}, shifts the $3\,^1\!D_2$ and $3\,^3\!D_1$ ionization energies by
0.34~MHz and 0.27~MHz, respectively.

Table~\ref{tab:energy} presents a summary of individual contributions to the ionization energy of
the $3\,^1\!D_2$ and $3\,^3\!D_J$ states of the $^4$He atom. Our theoretical values of the
ionization energies differ from the previous results of Morton {\em et al.}~\cite{morton:06:cjp} by
about 0.3~MHz, or $10\,\sigma$. The main reason for such a large deviation is the nonradiative
part of the $m\alpha^6$ correction described in the preceding paragraph.
Moreover, our final uncertainty is similar to that of of Morton
{\em et al.}, but in our case it comes from the higher-order $m\alpha^7$ contribution, which is
estimated by scaling the known result for the hydrogenic radiative corrections. Since we found that
in order $m\alpha^6$ the radiative correction is not dominant, we have to assume that a similar
situation can occur in the next order, so we estimate the uncertainty as 100\% of the radiative
effects. For the fine-structure and the singlet-triplet separation intervals, we keep the same
uncertainty as for the individual ionization energies, since we assume that the nonradiative
$m\alpha^7$ effects could contribute on the same level as the radiative ones.

\begin{table*}
\caption{$m\alpha^6$ corrections for ionization energies, in units of $10^{-3}\,m\alpha^6$.
Conversion factor to MHz is $0.018658054$. $S = 0,1$ denotes the spin of the reference state,
whereas $S\,' = 1-S$ denotes the opposite spin state
(triplet for singlet and vice versa). \label{tab:ma6}}
\begin{ruledtabular}
\begin{tabular}{lcdddd}
Contribution & Intermediate & \multicolumn{1}{c}{$3\,{}^1\!D$}
                                              & \multicolumn{3}{c}{$3\,{}^3\!D$}\\
             &  states symmetry  &           & \multicolumn{1}{c}{$J = 1$}
                                                   & \multicolumn{1}{c}{$J = 2$} &
                                                                    \multicolumn{1}{c}{$J = 3$} \\
\hline\\[-5pt]
$E_{Q}$      &               & 19.711         &  19.853      &  19.853      &  19.853 \\
$E_{H}$      &               & -0.006         &              &              &         \\
$E_{R1}$     &               &-10.667         & -13.220      & -13.220      & -13.220 \\
$E_{R2}$     &               & -0.098         &  -0.118      &  -0.118      &  -0.118 \\
$E_{LG}$     &               &  0.035         &              &                    \\
$E_{\rm fs, DK}$ &           &                &  -6.395      &   0.377      &   2.471  \\
$E_{\rm fs, amm}$ &          &                &  -0.050      &   0.051      &  -0.015 \\
$E_{\rm sec}$& $^{2S'+1}P$   & -0.018         &  -0.025      &              &          \\
             & $^{2S'+1}D$   & -1.156         &              &  -1.158      &          \\
             & $^{2S'+1}F$   & -0.057         &              &             &   -0.039  \\
             & $^{2S+1}S$    &                &   0.148      &              &          \\
             & $^{2S+1}P$    &                &  -0.021      &  -0.145      &          \\
             & $^{2S+1}D$    & -0.156\,(2)    &   0.705\,(8) &   0.233\,(2) &   0.313\,(6)  \\
             & $^{2S+1}F$    &                &              &  -0.056      &  -0.057  \\
             & $^{2S+1}G$    &                &              &              &  -0.013  \\
[5pt]
Total        &               &  7.589\,(2)    &   0.873\,(8) &   5.816\,(2) &   9.175\,(6)\\
Total(MHz)   &               &  0.142         &   0.016      &   0.109      &   0.171\\
\end{tabular}
\end{ruledtabular}
\end{table*}

\begin{table*}
\caption{Theoretical ionization energies of the $1s3d$ states of $^4$He, in MHz.
The values of fundamental constants used are $R_{\infty}c = 3\,289\,841\,960.355$~MHz,
$\alpha^{-1} = 137.035\,999\,139$, $ M/m = 7294.29954136$.
Uncertainties of fundamental constants do not influence the numerical results presented.
\label{tab:energy}}
\begin{ruledtabular}
\begin{tabular}{ldddd}
                     &        \multicolumn{1}{c}{$3^1\!D_2$}
                                            &  \multicolumn{1}{c}{$3{}^3\!D_1$}
                                                                   &  \multicolumn{1}{c}{$3{}^3\!D_2$}
                                                                                          &  \multicolumn{1}{c}{$3^3\!D_3$}
                                                                   \\
                                                                   \hline\\[-5pt]
$E^{(2,0)}$          & -365\,966\,841.606       & -366\,069\,330.717       & -366\,069\,330.717       & -366\,069\,330.717       \\
$E^{(2,1)}$          &      49\,946.656       &      50\,208.515       &      50\,208.515       &      50\,208.515       \\
$E^{(2,2+)}$         &        -13.886       &        -13.652       &        -13.652       &        -13.652       \\
$E^{(4,0)}$          &       -851.144       &        259.290       &      -1\,039.409       &      -1\,141.056       \\
$E^{(4,1)}$          &          0.154       &         -0.465       &          0.081       &          0.143       \\
$E^{(5,0)}$          &        -13.962       &        -15.707       &        -17.705       &        -16.413       \\
$E^{(5,1)}$          &         -0.004       &          0.003       &          0.004       &          0.004       \\
$E_{\rm MIX}$        &         24.967\,(5)  &          0.          &        -24.967\,(5)  &          0.          \\
$E^{(6,0)}$          &          0.142       &          0.016       &          0.109       &          0.171       \\
$E^{(7,0)}$          &          0.019\,(19) &          0.023\,(23) &          0.023\,(23) &          0.023\,(23) \\
$E_{\rm FNS}$        &         -0.008       &         -0.009       &         -0.009       &         -0.009       \\
Total theory         & -365\,917\,748.673\,(20) & -366\,018\,892.702\,(23) & -366\,020\,217.728\,(24) & -366\,020\,292.992\,(23) \\
Previous theory \cite{morton:06:cjp} &
                       -365\,917\,749.02\,(2)   & -366\,018\,892.97\,(2)   & -366\,020\,218.09\,(2)   & -366\,020\,293.41\,(2) \\
Difference           &          0.35\,(3)   &          0.27\,(3)   &          0.36\,(3)   &          0.42\,(3)\\
\end{tabular}
\end{ruledtabular}
\end{table*}

Tables~\ref{tab:fs} and \ref{tab:trans} present comparisons of theoretical predictions with
experimental results for the fine-structure intervals and various transition frequencies for the
$^4$He atom. The result for the $3\,^1\!D_2 $--$ 3\,^3\!D_1$ transition is obtained by combining
together four measurements \cite{Huang:18,rengelink:18,zheng:17,luo:16},
\begin{align}
  E(3^1D_2 - 3^3D_1) = E(3^1D_2 - 2^1S_0) + E(2^1S_0 - 2^3S_1)\nonumber \\ - E(2^3P_0 - 2^3S_1) - E(3^3D_1 - 2^3P_0)
\,.  \label{combined}
\end{align}

\begin{table}
\caption{Fine-structure energy differences of the $3{}^3\!D_J$ states of $^4$He, in MHz.
\label{tab:fs} }
\begin{ruledtabular}
\begin{tabular}{llll}
\multicolumn{1}{c}{$\nu_{32}$} & \multicolumn{1}{c}{$\nu_{21}$}  & \multicolumn{1}{c}{$\nu_{31}$}\\
\hline\\[-5pt]
%
  $-75.264\,(24)$ &  $-1325.026\,(24)$  &  $-1400.290\,(23)$ & This work\\
  $-75.32\,(2)  $ &  $-1325.12\,(2)  $  &  $-1400.44\,(2)  $ & theo. \cite{morton:06:cjp} \\
  $-76.15\,(30) $ &  $-1324.50\,(35) $  &  $-1400.65\,(37) $ &  exp. \cite{peterschmann:83}\\
  $-75.97\,(23) $ &                     &  $-1400.67\,(29) $ & exp.    \cite{Tam:75}\\
\end{tabular}
\end{ruledtabular}
\end{table}

\begin{table*}
  \caption{Comparison of different theoretical predictions with experimental results for various transition energies
  in $^4$He, in MHz.
  Theoretical ionization energies of the $n=2$ states in the column ``Present theory'' are taken from Ref.~\cite{pachucki:17}.
   \label{tab:trans}
}
\begin{center}
\begin{ruledtabular}
\begin{tabular}{r w{11.6} c w{11.6} w{2.6} w{11.6} w{2.6}}
                            & \centt{Experiment}
                            & \centt{Ref.}
                            & \centt{Present theory}
                            & \centt{Difference}
                            & \centt{Other theory}
                            & \centt{Difference}
                            \\
                            &
                            &
                            &
                            &  \centt{from experiment}
                            & \centt{\cite{morton:06:cjp}}
                            &  \centt{from experiment}
                            \\\hline
\\[-1.0ex]
\multicolumn{3}{l}{$3L'$--$2L$ transitions:}\\[0.5ex]
  $3^1D_2$--$2^1S_0$        & 594\,414\,291.803\,(13)
                            &\cite{Huang:18}
                            & 594\,414\,289.3\,(1.9)
                            &             2.5\,(1.9)
                            & 594\,414\,292.\,(5.)
                            &             0.\,(5.)
                            \\ [0.5ex]
  $3^3D_1$--$2^3S_1$        & 786\,823\,850.002\,(56) \,
                            &\cite{Dorrer:97}
                            & 786\,823\,848.7\,(1.3)\,
                            &             1.3\,(1.3)
                            & 786\,823\,845.\,(7.)
                            &             4.\,(7.)
                            \\ [0.5ex]
  $3^3D_1$--$2^3P_0$        & 510\,059\,755.352\,(28) \,
                            &\cite{luo:16}
                            & 510\,059\,754.2\,(0.7)\,
                            &             1.2\,(0.7)\,
                            & 510\,059\,749.\,(2.)
                            &             6.\,(2.)
                            \\ [0.5ex]
  $3^1D_2$--$2^1P_1$        & 448\,791\,399.113\,(268)\,
                            &\cite{luo2}
                            & 448\,791\,397.8\,(0.4)\,
                            &             1.3\,(0.5)
                            & 448\,791\,400.5\,(2)
                            &            -1.4\,(2)
                            \\  [1.5ex]
\multicolumn{3}{l}{$2L'$--$2L$ transitions:}\\[0.5ex]
  $2^3P_0$--$2^3S_1$        & 276\,764\,094.657\,2\,(14) \,
                            &\cite{zheng:17}
                            & 276\,764\,094.5\,(2.0)
                            &             0.2\,(2.0)
                            & 276\,764\,096.\,(7.)
                            &             2.\,(7.)
                            \\  [0.5ex]
  $2^1S_0$--$2^3S_1$        & 192\,510\,702.148\,72\,(20) \,
                            &\cite{rengelink:18}
                            & 192\,510\,703.4\,(0.8)
                            &            -1.3\,(0.8)
                            & 192\,510\,697.\,(9.)
                            &             5.\,(9.)
                            \\ [0.5ex]
  $2^1P_1$--$2^1S_0$        & 145\,622\,892.886\,(183) \,
                            &\cite{luo1,*luo1:erratum}
                            & 145\,622\,891.5\,(2.3)\,
                            &             1.4\,(2.3)\,
                            & 145\,622\,892.\,(5.)
                            &             0.\,(5.)
                            \\ [0.5ex]
  $2^1P_1$--$2^3S_1$        & 338\,133\,594.4\,(5) \,
                            & \cite{notermans:14}
                            & 338\,133\,594.9\,(1.4)
                            &            -0.5\,(2.2)
                            & 338\,133\,589.\,(7.)
                            &             5.\,(7.)
                            \\  [1.5ex]
\multicolumn{3}{l}{$3L'$--$3L$ transitions:}\\[0.5ex]
  $3^1D_2$--$3^3D_1$        & 101\,143.943\,(31) \,
                            & \cite{Huang:18,rengelink:18,zheng:17,luo:16}
                            & 101\,144.029\,(23)
                            &        0.086\,(37)
                            & 101\,143.95\,(3)
                            &        0.01\,(4)
                            \\
\end{tabular}
\end{ruledtabular}
\end{center}
\end{table*}

For the fine structure, we observe deviations of both sets of theoretical predictions, ours and
those of Morton {\em et al.}, from the experimental results on the level of 2 -- 3 of experimental
$\sigma$. The experiments are rather old and their accuracy is lower than what could be achievable
nowadays, so it is desirable to verify them before any definite conclusions are drawn.

The comparison of theory and experiment for transition frequencies presented in
Table~\ref{tab:trans} is quite surprising. We observe good agreement between theory and experiment
for all measured $2L'$--$2L$ transitions. For the $3D$--$2L$ intervals, however, all experimental
transition frequencies are about 1~MHz larger than the theoretical predictions. Since different
experimental results are supposed to be uncorrelated, a reason for the systematic discrepancy
should be on the theoretical side. An unaccounted-for contribution of 1~MHz could hardly come from
the $3D$ ionization energy since two independent calculations (ours and that of Drake and
co-workers \cite{morton:06:cjp}) agree on this level of accuracy. This would mean that an unknown,
nearly $L$-independent contribution of about 1~MHz is present for all $n=2$ ionization energies.
Assuming the standard $1/n^3$ scaling of QED effects, this implies an unknown contribution of
$10/n^3$~MHz for an arbitrary state.

Having in mind that theoretical energies of the $n=2$ states of helium have been independently
checked on the level of the $m\alpha^5$ effects \cite{yerokhin:10:helike,drake:05:springer},
possible sources of unaccounted contributions could be a mistake in the evaluation of the
$m\alpha^6$ corrections or an underestimation of $m\alpha^7$ effects. The latter possibility will
be checked when our ongoing project of calculating all $m\alpha^7$ effects to the $2\,^3\!S$ and
$2\,^3\!P$ ionization energies \cite{yerokhin:18:betherel} is completed.

On the experimental side, it is desirable to conduct more measurements of transitions between
states from different shells, as this will allow to confirm and study further the systematic
deviation of experimental results from theoretical predictions.

In summary, we performed detailed calculations of ionization energies of the $1s3d$ states in the
$^4$He atom, including the complete evaluation of the $m\alpha^6$ QED effects.  The nonradiative
$m\alpha^6$ corrections, which have not been accounted for in the previous calculations, turned out
to be much larger than previously anticipated, shifting the theoretical predictions by about
10$\,\sigma$. However, this was not sufficient to explain the previously reported
systematic discrepancies between the theoretical and experimental results for the $3D$--$2L$
transitions. These discrepancies could possibly indicate the presence of some unaccounted-for
contributions of order $m\alpha^6$ or underestimation of higher-order effects.

\begin{acknowledgments}
This work was supported by the National Science Center (Poland) Grant No. 2017/27/B/ST2/02459.
V.A.Y. acknowledges support by the Ministry of Education and Science of the Russian Federation
Grant No. 3.5397.2017/6.7. V.P. acknowledges support from the Czech Science Foundation - GA\v{C}R
(Grant No. P209/18-00918S).
\end{acknowledgments}

\appendix

\section{Wave functions in Cartesian coordinates}
Since we use the explicitly correlated basis functions, it is convenient to represent the angular
part of the wave function in Cartesian coordinates. In this section we list the explicit
expressions for wave functions of symmetries relevant for this work.
We denote by $(...)^{(n)}$ the traceless
and symmetric rank-$n$ tensor and $\vec{R} \equiv \vec{r}_1\times \vec{r}_2$.

The $L = 0$ wave function of a definite exchange symmetry is of the form
\begin{align}
\phi\left(^{1,3}\!S^e\right) = F \pm (1\leftrightarrow2)\,.
\end{align}
where $F$ is a scalar function of $r_1$, $r_2$ and $r \equiv |\vec{r}_1-\vec{r}_2|$, the upper sign
corresponds to the singlet and the lower sign, to the triplet state.

The $L = 1$ odd and even wave functions are:
\begin{align}
\vec{\phi}\left(^{1,3}\!P^o\right) &\ = \vec{r}_1 \, F \pm (1\leftrightarrow2)\,, \\
\vec{\phi}\left(^{1,3}\!P^e\right) &\ = \vec{R} \, F \pm (1\leftrightarrow2)\,.
\end{align}

The $L = 2$ odd and even wave functions are:
\begin{align}
\phi^{ij}\left(^{1,3}\!D^o\right) &\ = \bigl( r_1^i R^j + r_1^j  R^i\bigr)\, F
\pm (1\leftrightarrow2)\,,
 \\\
\phi^{ij}\left(^{1,3}\!D^e\right) &\ = \big(r_1^ir_1^j\big)^{(2)}\,F + \big(r_1^ir_2^j\big)^{(2)}\,G
\pm (1\leftrightarrow2)\,,
\end{align}
where
\begin{align}
\big(r_1^ir_1^j\big)^{(2)} &\ = r_1^jr_1^j - \frac{1}{3}\,\delta^{ij}\,r_1^2\,,\\
\big(r_1^ir_2^j\big)^{(2)} &\ = \frac12\Big( r_1^jr_2^j + r_2^jr_1^j -\frac{2}{3}\, \delta^{ij}\, \vec{r}_1\cdot\vec{r}_2\Big)\,\,.
\end{align}

The $L = 3$ odd and even wave functions are:
\begin{align}
\phi^{ijk}\left(^{1,3}\!F^o\right) & = \big(r_1^ir_1^jr_1^k\big)^{(3)}\,F + \big(r_1^ir_1^jr_2^k\big)^{(3)}\,G
\pm (1\leftrightarrow2)\,,
  \\
\phi^{ijk}\left(^{1,3}\!F^e\right) & = \big(r_1^ir_1^jR^k\big)^{(3)}\,F +
\big(r_1^ir_2^jR^k\big)^{(3)}\,G \pm (1\leftrightarrow2)\,,
\end{align}
where
\begin{widetext}
\begin{align}
\big(r_1^ir_1^jr_1^k\big)^{(3)} = &\
   r_1^ir_1^jr_1^k - \frac{r_1^2}5\,\left(\delta^{ij}r_1^k+ \delta^{ik}r_1^j+ \delta^{jk}r_1^i
\right)\,,
  \\
\big(r_1^ir_1^jr_2^k\big)^{(3)} = &\
\frac13 \, \Bigl[ r_1^i r_1^j r_2^k + r_1^i r_2^j r_1^k + r_2^i r_1^j r_1^k
-\frac{r_1^2}{5}  \Big( \delta^{ij}r_2^k + \delta^{ik}r_2^j + \delta^{jk}r_2^i\Big)
-\frac{2\, \vec{r}_1\cdot\vec{r}_2}{5}  \Big( \delta^{ij}r_1^k + \delta^{ik}r_1^j +
\delta^{jk}r_1^i\Big)
 \Bigr]\,,
 \\
\big(r_1^ir_1^jR^k\big)^{(3)} = &\ \frac13\,\Big[ r_1^ir_1^jR^k + r_1^iR^jr_1^k + R^ir_1^jr_1^k
 -\frac{r_1^2}{5}  \Big( \delta^{ij}R^k + \delta^{ik}R^j + \delta^{jk}R^i\Big)\Big]\,,
  \\
\big(r_1^ir_2^jR^k\big)^{(3)} = &\ \frac16\,\Big[ r_1^ir_2^jR^k + r_1^iR^jr_2^k + R^ir_1^jr_2^k
                                             +  r_2^ir_1^jR^k + r_2^iR^jr_1^k + R^ir_2^jr_1^k
 -\frac{2\,\vec{r}_1\cdot\vec{r}_2}{5}  \Big( \delta^{ij}R^k + \delta^{ik}R^j + \delta^{jk}R^i\Big)\Big]\,.
\end{align}

The $L = 4$ even wave function is:
\begin{align}
\phi^{ijkl}\left(^{1,3}G^e\right) &\ = \big(r_1^i r_1^j r_1^k r_1^l\big)^{(4)}\,F
                                       + \big(r_1^i r_1^j r_1^k r_2^l\big)^{(4)}\,G
                                       + \big(r_1^i r_1^j r_2^k r_2^l\big)^{(4)}\,H
\pm (1\leftrightarrow2)\,,
\end{align}
where
\begin{align}
\big(r_1^i r_1^j r_1^k r_1^l\big)^{(4)} = &\  r_1^i r_1^j r_1^k r_1^l -\frac{r_1^2}{8}\Big(
\delta^{ij} r_1^k r_1^l
+ \delta^{ik} r_1^j r_1^l
+ \delta^{il} r_1^j r_1^k
+ \delta^{jk} r_1^i r_1^l
+ \delta^{jl} r_1^i r_1^k
+ \delta^{kl} r_1^i r_1^j
\Big)\,,\\
\big(r_1^i r_1^j r_1^k r_2^l\big)^{(4)} = &\  \frac14\bigg[
r_1^i r_1^j r_1^k r_2^l
+ r_1^i r_1^j r_2^k r_1^l
+ r_1^i r_2^j r_1^k r_1^l
+ r_2^i r_1^j r_1^k r_1^l
  \nonumber \\ &
-\frac18\Big(
\delta^{ij} S^{kl}
+\delta^{ik} S^{jl}
+\delta^{il} S^{jk}
+\delta^{jk} S^{il}
+\delta^{jl} S^{ik}
+\delta^{kl} S^{ij}
\Big)\bigg]\,,\\
S^{kl} = &\  r_1^2 \big(r_1^kr_2^l+r_2^kr_1^l\big) +
2\, \vec{r}_1 \cdot \vec{r}_2\, r_1^k r_1^l\,,\\[1ex]
\big(r_1^i r_1^j r_2^k r_2^l\big)^{(4)} =  &\  \frac16\bigg[
  r_1^i r_1^j r_2^k r_2^l
+ r_1^i r_2^j r_1^k r_2^l
+ r_1^i r_2^j r_2^k r_1^l
+ r_2^i r_1^j r_1^k r_2^l
+ r_2^i r_1^j r_2^k r_1^l
+ r_2^i r_2^j r_1^k r_1^l
  \nonumber \\ &
-\frac18\Big(
\delta^{ij} P^{kl}
+\delta^{ik} P^{jl}
+\delta^{il} P^{jk}
+\delta^{jk} P^{il}
+\delta^{jl} P^{ik}
+\delta^{kl} P^{ij}
\Big)\bigg]\,,\\
P^{kl} = &\ r_1^2 r_2^kr_2^l + r_2^2 r_1^kr_1^l
+ 2\, \vec{r}_1 \cdot \vec{r}_2\, \big(r_1^k r_2^l+r_2^k r_1^l\big)\,.
\end{align}
\end{widetext}

\section{Tensor decomposition in Cartesian coordinates}
\label{app:tensor}

In order to perform the angular-momentum algebra in Cartesian coordinates, one requires
decompositions of products of various operators into traceless and symmetric tensors. First, we
decompose the product of a traceless and symmetric tensor $D^{ij}$ and an arbitrary vector $Q^k$,
\begin{align}
  D^{ij}\,Q^k =&\ T^{ijk} + \epsilon^{ikl}\,T^{lj} + \epsilon^{jkl}\,T^{li}
  \nonumber \\&
  +\delta^{ik}\,T^j +\delta^{jk}\,T^i -\frac{2}{3}\,\delta^{ij}\,T^k\,,
\end{align}
where
\begin{align}
    T^{ijk} =&\ (D^{ij}\,Q^k)^{(3)}\,, \\
    T^{ij} =&\ \frac{1}{6}\,\bigl(\epsilon^{jkl} D^{ik}\,Q^l + \epsilon^{ikl} D^{jk}\,Q^l\bigr)\,, \\
    T^i =&\ \frac{3}{10}\,D^{ij} Q^j\,.
\end{align}
This decomposition was used in calculations of various second-order matrix elements and in the
evaluation of the Bethe logarithm. In the latter case, with $Q^k = p^k$, we obtain
\begin{align}
  \Big< D^{ij} p^k \, \hat R\, p^k D^{ij}\Big> =&\
  \Big< D^{ij}\,p^k\,\hat R\,(D^{ij}\,p^k)^+\Big>
  \nonumber \\
  = &\
\Big< (D^{ij}\,p^k)^{(3)}\,\hat R\,\left.(D^{ij}\,p^k)^{{(3)}}\right.^+ \Big>
\nonumber \\ &
+ 6\,\Big< T^{ij}\,\hat R\, \left. T^{ij} \right.^+\Big>
+\frac{20}{3}\,\Big< T^i\,\hat R\,\left. T^{i}\right.^+ \Big> \,,
\end{align}
where $\hat R = 1/(E-H)$ and ``$+$'' denotes the Hermitian conjugate.

The second decomposition we need is that of the product of two traceless and symmetric tensors
$D^{ij}$ and $Q^{ij}$,
\begin{widetext}
\begin{align}
D^{ij}\,Q^{kl} =&\  T^{ijkl} + \epsilon^{ika}T^{jal} + \epsilon^{jka}T^{ial} + \epsilon^{ila}T^{jak} + \epsilon^{jla}T^{iak} +
\delta^{ik}T^{jl}+\delta^{il}T^{jk} + \delta^{jk}T^{il}+\delta^{jl}T^{ik}
\nonumber \\ &\ - \frac{4}{3}\,\delta^{ij}T^{kl}-\frac{4}{3}\,\delta^{kl} T^{ij} +
                  T^a\,\big(\epsilon^{ika}\delta^{jl} + \epsilon^{ila}\delta^{jk} +
                  \epsilon^{jka}\delta^{il} + \epsilon^{jla}\delta^{ik}\big) +
                  T\,\big(\delta^{ik}\delta^{jl} + \delta^{il}\,\delta^{jk}-\frac{2}{3}\,\delta^{ij}\delta^{kl}\big)\,,
\end{align}
\end{widetext}
where
\begin{align}
  T^{ijkl} =&\ (D^{ij}\,Q^{kl})^{(4)}\,,\\
  T^{jbl} =&\ \frac{1}{4}\,(\epsilon^{ikb}\,D^{ij}\,Q^{kl})^{(3)}\,,\\
  T^{jl} =&\ \frac{3}{7}\,(D^{ij}\,Q^{il})^{(2)}\,,\\
  T^b =&\ \frac{1}{10}\,\epsilon^{jlb}\,D^{ij}\,Q^{il}\,,\\
  T =&\ \frac{1}{10}\,D^{ij}\,Q^{ij}\,.
  \end{align}

\section{Spin-angular representation of D-states}
Let $\vec S$ be the angular momentum operator for $S=1$ that satisfies the commutator relation
\begin{equation}
  [S^i,S^j] = i\,\epsilon^{ijk}\,S^k\,,
\end{equation}
then in the fundamental representation
\begin{equation}
S^i\,S^j\,S^k = \frac{i}{2}\,\epsilon^{ijk}\,\vec S^{\,2}+\delta^{jk}\,S^i + i\,\epsilon^{ika}\,S^j\,S^a
\end{equation}
and
\begin{align} \label{tr1}
  {\rm Tr}\,S^i\,S^j =&\ 2\,\delta^{ij}\,, \\
  {\rm Tr}\,S^i\,S^j\,S^k =&\ i\,\epsilon^{ijk}\,, \\
  {\rm Tr}\,S^i\,S^j\,S^k\,S^l =&\ \delta^{ij}\,\delta^{kl} + \delta^{jk}\,\delta^{il}\,.
\end{align}
Assuming the explicit representation of the spin operator in terms of Pauli matrices, $\vec S =
(\vec\sigma_1+\vec\sigma_2)/2$, we obtain the following identities,
\begin{widetext}
\begin{align}
  \frac{1}{5}\,\sum_{M} |^1D_{2M}\rangle\,\langle ^1D_{2M}| =&\
  |^1D^{ij}\rangle\langle ^1D^{ij}|\,\Bigl(1-\frac{\vec S^{\,2}}{2}\Bigr) \,,\\
  \frac{1}{3}\,\sum_{M} |^3D_{1M}\rangle\,\langle ^3D_{1M}| =&\
  |^3D^{ik}\rangle\langle ^3D^{jk}|\,\Bigl(\delta^{ij}\,\frac{\vec S^{\,2}}{2} -S^j\,S^i\Bigr) \,,\\
  \frac{1}{5}\,\sum_{M} |^3D_{2M}\rangle\,\langle ^3D_{2M}| =&\
  |^3D^{ik}\rangle\langle ^3D^{jk}|\,\Bigl(-\frac{1}{3}\,\delta^{ij}\,\frac{\vec S^{\,2}}{2}
  +\frac{2}{3}\,S^i\,S^j +\frac{1}{3}\,S^j\,S^i\Bigr) \,,\\
  \frac{1}{7}\,\sum_{M} |^3D_{3M}\rangle\,\langle ^3D_{3M}| =&\
  |^3D^{ik}\rangle\langle ^3D^{jk}|\,\Bigl(\frac{11}{21}\,\delta^{ij}\,\frac{\vec S^{\,2}}{2}
  -\frac{10}{21}\,\,S^i\,S^j +\frac{4}{21}\,S^j\,S^i\Bigr)\,.
  \label{tr2}
  \end{align}

\section{Explicit formulas for the second-order corrections}
\label{app:sec}

In this section we present explicit calculation formulas for the second-order corrections, for the
singlet ($S = 0$) and triplet ($S = 1$) reference states. For each reference state, there are four
different symmetries of intermediate states contributing, with rational weight factors that are
determined by the angular-momentum algebra method illustrated in the previous sections. The results
are as follows. { For ${S = 1}$ and ${ J = 1}$},
\begin{align}
E_{\rm sec}(\,^3\!D_1) =&\   E(\,^3\!D_1| \,^1\!P^e) + E(\,^3\!D_1| \,^3\!S) + E(\,^3\!D_1| \,^3\!P^e) + E(\,^3\!D_1| \,^3\!D)\,,\\
E(\,^3\!D_1| \,^1\!P^e) =&\ \sum_n\frac{1}{E-E_n}\,\Big< {}\,^3\!D^{ij}\Big|\im\,Q_C^j\Big|{}\,^1\!P^{i}_n\Big>^2\,, \\
E(\,^3\!D_1| \,^3\!S) =&\ \frac43\,\sum_n \frac1{E-E_n}\,\Big< {}\,^3\!D^{ij}\Big| Q_D^{ij}\Big| \,^3\!S_n\Big>^2\,,\\
E(\,^3\!D_1| \,^3\!P^e) =&\ \frac12\,\sum_n \frac1{E-E_n}\,\Big< {}\,^3\!D^{kj}\Big|\im\,\delta^{ki}\, Q_B^j
                   - 2\,\epsilon^{jli} Q_D^{kl}\Big| {}\,^3\!P^i_n\Big>^2 \,,\\
E(\,^3\!D_1| \,^3\!D) =&\ \left. \sum_n \right.^{\prime\prime} \frac1{E-E_n}\, \Big< {}\,^3\!D^{ik}\Big| \delta^{kj}\,Q_{A{\rm reg}}
                  + \im\,\epsilon^{klj}Q_B^{l} -2\,Q_D^{kj}\Big| {}\,^3\!D^{ij}_n\Big>^2 \,.
\end{align}

{ For ${S = 1}$ and ${ J = 2}$},
\begin{align}
E_{\rm sec}(\,^3\!D_2) =&\ E(\,^3\!D_2| \,^1\!D) + E(\,^3\!D_2|\,^3\!P^e) + E(\,^3\!D_2|\,^3\!D) + E(\,^3\!D_2|\,^3\!F^e)\,,\\
E(\,^3\!D_2| \,^1\!D) = &\
  \frac{2}{3}\,\left. \sum_n \right.^{\prime\prime}\frac{1}{E-E_n}\,\Big< {}\,^3\!D^{ik}\Big|\epsilon^{klj}\im\,Q_C^l\Big|{}\,^1\!D^{ij}_n\Big>^2\,,  \\
E(\,^3\!D_2| \,^3\!P^e) =&\ \frac1{10}\,\sum_n \frac1{E-E_n}\, \Big< {}\,^3\!D^{kj}\Big|3\,\im\,\delta^{ki}\, Q_B^j
                   + 2\,\epsilon^{jli} Q_D^{kl}\Big| {}\,^3\!P^i_n\Big>^2 \,,\\
E(\,^3\!D_2|\,^3\!D) =&\ \left. \sum_n \right.^{\prime\prime} \frac1{E-E_n}\,\Big< {}\,^3\!D^{ik}\Big| \delta^{jk}\,Q_{A{\rm reg}}
                 + \frac{\im}{3}\,\epsilon^{klj}Q_B^{l}
                 + 2\,Q_D^{kj}\Big| {}\,^3\!D^{ij}_n\Big>^2 \,,\\
E(\,^3\!D_2|\,^3\!F^e) =&\ \frac23\,\sum_n \frac1{E-E_n}\, \Big< {}\,^3\!D^{ia}\Big|\im\,\delta^{ja}\, Q_B^k
                  - 2\,\epsilon^{abj} Q_D^{bk}\Big| \,{}\,^3\!F^{ijk}_n\Big>^2 \,.
\end{align}

{ For ${S = 1}$ and ${ J = 3}$},
\begin{align}
  E_{\rm sec}(\,^3\!D_3) =&\ E(\,^3\!D_3| \,^1\!F^e) + E(\,^3\!D_3| \,^3\!D) + E(\,^3\!D_3| \,^3\!F^e) + E(\,^3\!D_3|\,^3\!G)\,,\\
  E(\,^3\!D_3| \,^1\!F^e) =&\  \frac{5}{7}\,\sum_n\frac{1}{E-E_n}\,\Big< {}\,^3\!D^{ij}\Big|\im\,Q_C^k\Big|{}\,^1\!F^{ijk}_n\Big>^2\,, \\
  E(\,^3\!D_3| \,^3\!D) =&\ \left. \sum_n \right.^{\prime\prime} \frac1{E-E_n}\, \Big< {}\,^3\!D^{ik}\Big| \delta^{jk}\,Q_{A{\rm reg}}
                   - \frac23\,\im\,\epsilon^{klj}Q_B^{l}
                   -\frac47\,Q_D^{kj}\Big| {}\,^3\!D^{ij}_n\Big>^2 \,,\\
E(\,^3\!D_3| \,^3\!F^e) =&\ \frac{20}{21}\,\sum_n \frac1{E-E_n}\,\Big< {}\,^3\!D^{ia}\Big|\im\,\delta^{ja}\, Q_B^k
                   + \epsilon^{abj} Q_D^{bk}\Big| \,{}\,^3\!F^{ijk}_n\Big>^2 \,,\\
E(\,^3\!D_3|\,^3\!G) =&\ \frac{20}{7}\,\sum_n \frac1{E-E_n}\,\Big< \,^3\!D^{ij}\Big|Q_D^{kl}\Big| \,\,^3\!G^{ijkl}_n\Big>^2 \,.
\end{align}

{ For ${S = 0}$ and ${ J = 2}$},
\begin{align}
  E_{\rm sec}(\,^1\!D_2) =&\ E(\,^1\!D_2| \,^1\!D) + E(\,^1\!D_2|\,^3\!P^e) + E(\,^1\!D_2|\,^3\!D) + E(\,^1\!D_2|\,^3\!F^e)\,,\\
  E(\,^1\!D_2| \,^1\!D)  =&\ \left. \sum_n \right.^{\prime\prime}\frac{1}{E-E_n}\,\Big< \,^1\!D^{ij}\Big|Q_{A{\rm reg}}\Big|\,^1\!D^{ij}_n\Big>^2 \,,  \\
  E(\,^1\!D_2|\,^3\!P^e) =&\ \frac{3}{5}\,\sum_n\frac{1}{E-E_n}\,\Big< \,^1\!D^{ij}\Big|\im Q_C^j\Big|\,^3\!P^{i}_n\Big>^2 \,, \\
  E(\,^1\!D_2|\,^3\!D) =&\ \frac{2}{3}\,\left. \sum_n \right.^{\prime\prime}\frac{1}{E-E_n}\,\Big< \,^1\!D^{ik}\Big|\im\epsilon^{klj}Q_C^l\Big|\,^3\!D^{ij}_n\Big>^2\,, \\
  E(\,^1\!D_2|\,^3\!F^e) =&\ \sum_n\frac{1}{E-E_n}\,\Big< \,^1\!D^{ij}\Big|\im Q_C^k\Big|\,^3\!F^{ijk}_n\Big>^2\,.
\end{align}
In the formulas above, the double prime on the sum means that the singlet and triplet $3D$ states
are excluded from the summation over the spectrum.

\end{widetext}


\end{document}